\documentclass[prb,twocolumn,superscriptaddress,
longbibliography,aps]{revtex4-2}

\usepackage{graphicx}
\usepackage{bm}
\usepackage{color}
\usepackage{epstopdf}
\usepackage{amsmath}
\usepackage{amssymb}
\usepackage{epstopdf}

\usepackage{xfrac} 

\usepackage[urlcolor=blue,colorlinks=true,citecolor=blue,linkcolor=blue,pdfstartview={FitH},bookmarks=false]{hyperref}

\graphicspath{{fig/}{./fig/}{.}}

\begin{document}

\title{Low temperature phase of AuSn$_{4}$ induced by the van der Waals interactions}

\author{Shivam Yadav}
\email[e-mail: ]{shivam.yadav@ifj.edu.pl}
\affiliation{Institute of Nuclear Physics, Polish Academy of Sciences, ul. E. Radzikowskiego 152, PL-31342 Krak\'{o}w, Poland}

\author{Sajid Sekh}
\email[e-mail: ]{sajid.sekh@ifj.edu.pl}
\affiliation{Institute of Nuclear Physics, Polish Academy of Sciences, ul. E. Radzikowskiego 152, PL-31342 Krak\'{o}w, Poland}

\author{Andrzej Ptok}
\email[e-mail: ]{aptok@mmj.pl}
\affiliation{Institute of Nuclear Physics, Polish Academy of Sciences, ul. E. Radzikowskiego 152, PL-31342 Krak\'{o}w, Poland}

\date{\today}

\begin{abstract}
AuSn$_{4}$ is the example of an compound that exhibits topological properties. 
Recent XRD measurements reveal an ambiguous nature of the crystal structure, as it can be realized with either Aea2 or Ccca symmetry. 
Motivated by this, we analyze the dynamical stability of the compound. 
We discuss the role of van der Waals (vdW) corrections within the ab initio calculation.
Interestingly, our main result indicates that AuSn$_{4}$ can be unstable with both Aea2 and Ccca symmetries, due to the soft modes in the phonon spectra. 
From the soft mode analyses we find dynamically stable Pc structure.
This structure has always smaller energy than Aea2 or Ccca crystal and it stays independent of the vdW correction.
We also show that the comparison of theoretical electronic properties with the experimental ARPES measurements.
Our findings may be valuable in the future investigations of AuSn$_{4}$-like compounds.
\end{abstract}

\maketitle

\section{Introduction}

Experimental observation of the Dirac surface state in the topological insulators~\cite{zhang.liu.09,hsieh.xia.09,kuroda.arita.10,alpichshev.analytis.10} has started a period of intensive study of topological systems~\cite{hasan.kane.10,qi.zhang.11,armitage.mele.18,klemenz.lei.19,tokura.yasuda.19,lv.qian.21}.
One of the sources of topological properties is the spin--orbit coupling~\cite{soumyanarayanan.reyren.16}.
However, equally important is the symmetry realized by the studied system~\cite{song.zhang.18,chang.wieder.18,yu.wu.19,elcoro.song.20,li.fu.21,wu.tang.21,hirschmann.moritz.21,tang.wan.22}.

In this context, the case of AuSn$_{4}$ is fascinating.
This compound is reported to be crystallized with one of the two space group symmetries: Aea2~\cite{kubiak.81,kubiak.wolcyrz.85,okamoto.93,shen.kuo.20} or Ccca~\cite{zavalij.zribi.02}. 
Both space groups yield indistinguishable Bragg positions within the powder X-ray diffraction (XRD) measurements~\cite{sharma.gurjar.23,karn.sharma.22}.
However, the XRD measurements of Au$_{1-x}$Ni$_{x}$Sn$_{4}$ ($x=0.25$ and $0.5$) indicates the realization of the Ccca phase~\cite{zavalij.zribi.02}. 
A similar problem with XRD measurements was reported for PdSn$_{4}$~\cite{karn.sharma.23}, while the initial study suggested the realization of Ccca~\cite{nylen.garcia.04} phase.
In the case of PtSn$_{4}$, the structure was recognized with Ccca~\cite{kunnen.niepmann.00,mun.ko.12} phase.

The $\eta$-phase structure of AuSn$_{4}$ is stable within the gold-tin series~\cite{okamoto.93}.
The normal state low-temperature magnetoconductivity is well described by the weak-antilocalization (WAL) transport formula~\cite{shen.kuo.20}, which strongly supports the concept of the surface electrons with Dirac-node states. 
The bulk AuSn$_{4}$ exhibits two-dimensional (2D) superconductivity~\cite{shen.kuo.20}. 
In this context, the experimental results hint towards the type-II superconductivity in AuSn$_{4}$, with $T_{c}$ around \mbox{$2.3$--$2.65$~K~\cite{shen.kuo.20,sharma.poonamrani.22,sharma.gurjar.23,herrera.wu.23}.}
Moreover, the superconducting state exhibits a two-fold anisotropy in this system~\cite{sharma.gurjar.23}.

Previous theoretical study~\cite{herrera.wu.23}, independent of the realized space groups, shows a combination of intricate band structure and three-dimensional (3D) Fermi surface. 
Whereas, a direct estimation of the topological invariant $\mathbb{Z}_{2}$ reveals a distinction between the Aea2 and Ccca phases~\cite{karn.sharma.22}. 
Nevertheless, a proper study of the topological properties demands true crystal symmetry of AuSn$_{4}$, which motivates this work.

AuSn$_{4}$ is a layered compound, which suggests significant role of the van der Waals (vdW) correction~\cite{herrera.wu.23}.
During our investigation, we preform systematical study of the effects of absence and presence of the vdW interactions on the phonon spectra.
We show that the phonon spectra of AuSn$_{4}$ contains soft modes for both Aea2 and Ccca symmetries in the case of DFT-D3--like corrections.
Such soft mode are absent in absence of the vdW correction, or when vdW interactions are included within the nonlocal vdW-DF functionals~\cite{dion.rydberg.04}.
From analyses of the phonon soft mode we found new Pc stable structure of AuSn$_{4}$.
Such symmetry possess smaller energy than Aea2 or Ccca, independent of the phonon soft mode.
This clearly suggests stabilization of the AuSn$_{4}$ with another symmetry in low temperatures as we initially expected.
Finally, we discuss the electronic properties considering the aforementioned symmetries, and compare it with the experimental ARPES data.

The paper is organized as follows. 
First, we provide the details of the computational techniques in Sec.~\ref{sec.theo}. 
Next, we present study of dynamical stability of Aea2 and Ccca structures in Sec.~\ref{sec.init_sym}.
This starts with a discussion of the crystal structure in~\ref{sec.crys}, which is followed by the analysis of lattice dynamics in Sec.~\ref{sec.dynam}. 
Since the soft mode analysis predicts a new symmetry, we investigate the physical properties of system with Pc symmetry in Sec.~\ref{sec.pc_sym}.
We start with the system stability discussion in Sec.~\ref{sec.ph7}, while electronic properties and surface states are described in Sec.~\ref{sec.ele}.
Our conclusions are summarized in Sec.~\ref{sec.sum}.


\section{Computational details}
\label{sec.theo}

The first-principles (DFT) calculations are performed using the projector augmented-wave (PAW) potentials~\cite{bloch.94} implemented in the Vienna {\it Ab initio} Simulation Package (VASP)~\cite{kresse.hafner.94,kresse.furthmuller.96,kresse.joubert.99} code (version 6.4.3).
The calculations are made within the Perdew--Burke--Ernzerhof (PBE) pseudopotentials~\cite{perdew.burke.96}, with the valence electron configuration $5d^{10}6s^{1}$ and $5s^{2}p^{2}$ for Au and Sn, respectively.
Note that, the calculation includes the effect of spin-orbit interaction. 
We set the energy cut-off of the plane-wave expansion to $400$~eV. 
The summation over the reciprocal space is realized with a $12 \times 12 \times 7$ {\bf k}-grid (for structures with longest lattice constant along $z$) in the Monkhorst--Pack scheme~\cite{monkhorst.pack.76}. 
First, we optimize the lattice vectors, along with the position of the atoms. 
As a break of the optimization loop, we set the energy differences to $10^{-6}$~eV and $10^{-8}$~eV for ionic and electronic degrees of freedom, respectively. 
The symmetry of the optimized system is analyzed using {\sc FindSym}~\cite{stoke.hatch.05} and {\sc SpgLib}~\cite{togo.tanaka.18}, while the momentum space analysis is done with using {\sc SeeK-path} tools~\cite{hinuma.pizzi.17}.

{\it vdW interactions.}---
Due to the layered structure of AuSn$_{4}$, and its similarity to the previous study~\cite{herrera.wu.23}, we implement the vdW correction.
In our investigation, we also study the role of the vdW correction on the obtained results.
Thus, we compare results obtained {\it (i)} without vdW term, {\it (ii)} with vdW corrections, and {\it (iii)} with nonlocal vdW-DF functionals.

In the second group, we compare vdW correction within: 
DFT-D3 method of Grimme with zero-damping function~\cite{grimme.antony.10}, 
DFT-D3 method with Becke--Johnson damping function~\cite{grimme.ehrlich.11}, 
and Tkatchenko--Scheffler method with iterative Hirshfeld partitioning~\cite{bucko.lebegue.13,bucko.legegue.14}.
In this class of vdW correction, the additional dispersion correction term is added to the conventional Kohn--Sham energy.
Dispersion correction is added to the total energy, potential, interatomic forces or stress tensor which naturally affects, e.g., lattice dynamics.

The third group, the nonlocal vdW-DF functionals~\cite{klimes.bowler.11}, were originally proposed by Dion {\it et al.}~\cite{dion.rydberg.04}.
Here, the semilocal exchange-correlation functional that is augmented with a nonlocal correlation functional that approximately accounts for dispersion interactions.
Such corrections are also included in the calculation of total energies and forces.
In our calculation we applied: optPBE-vdW of Klime\v{s} {\it et al.} and optB88-vdW of Klime\v{s} {\it et al.}~\cite{klime.bowler.10}.

{\it Lattice dynamics.}---
The phonon calculations are realized within the {\it Parlinski--Li--Kawazoe} method~\cite{parlinski.li.97} implemented in Phonopy package~\cite{togo.tanaka.15}. 
In our calculation, the force constants were obtained from first-principles calculations of the Hellmann--Feynman forces by VASP -- this is used to build a dynamical matrix of the crystal. 
We compute the phonon frequencies by diagonalizing the dynamical matrix. 
Our calculations are performed using the supercell technique, related to the $2 \times 2 \times 1$ conventional cell with 16 formula units, and reduced $4 \times 4 \times 4$ {\bf k}-grid.

{\it Electronic surface states.}---
Finally, we study the topological properties using the tight-binding model in the maximally localized Wannier orbitals basis~\cite{marzari.vanderbilt.97,souza.marzari.01,marzari.mostofi.12}. 
This model was constructed from the exact DFT calculations in a primitive unit cell, with $8 \times 8 \times 6$ $\Gamma$-centered {\bf k}--point grid, using the {\sc Wannier90} software~\cite{pizzi.vitale.20}. 
The electronic surface states were calculated using the surface Green's function technique for a semi-infinite system~\cite{sancho.sancho.85}, implemented in {\sc WannierTools}~\cite{wu.zhang.18}.


\begin{table}[!t]
\caption{
\label{tab.latt_param}
Comparison of the lattice parameters of compounds with different symmetries (as labeled).
The results are obtained for the conventional unit cell related to the Aea2 convention (i.e. for lattice parameters $a < b < c$).
}
\begin{ruledtabular}
\begin{tabular}{ccccc}
 & & a (\AA) & b (\AA) & c (\AA) \\
\hline 
\hline 
\multicolumn{5}{c}{Experimental values} \\
\hline
Aea2 (SG:41) & Ref.~\cite{kubiak.wolcyrz.84} & 6.5124 & 6.5162 & 11.7065 \\
Aea2 (SG:41) & Ref.~\cite{sharma.gurjar.23} & 6.5216 & 6.5251 & 11.7301 \\
Aea2 (SG:41) & Ref.~\cite{karn.sharma.22} & 6.5036 & 6.5186 & 11.7173 \\
Ccca (SG:68) & Ref.~\cite{sharma.gurjar.23} & 6.4960 & 6.5471 & 11.7566 \\
Ccca (SG:68) & Ref.~\cite{karn.sharma.22} & 6.6582 & 6.6824 & 11.8716 \\
\hline
\hline
\multicolumn{5}{c}{This work} \\
\hline
\multicolumn{5}{c}{w/o vdW} \\
\hline
Aea2 (SG:41) & & 6.6132 & 6.6358 & 11.9324 \\
Ccca (SG:68) & & 6.6130 & 6.6360 & 11.9336 \\
\hline
\multicolumn{5}{c}{vdW, DFT-D3 method of Grimme} \\
\hline
Aea2 (SG:41) & & 6.5448 & 6.5506 & 11.7768 \\
Ccca (SG:68) & & 6.5450 & 6.5506 & 11.7768 \\\hline
\multicolumn{5}{c}{vdW, DFT-D3 method with Becke--Johnson damping} \\
\hline
Aea2 (SG:41) & & 6.5146 & 6.5405 & 11.7396 \\
Ccca (SG:68) & & 6.5145 & 6.5404 & 11.7403 \\
\hline
\multicolumn{5}{c}{vdW, Tkatchenko--Scheffler method} \\
\hline
Aea2 (SG:41) & & 6.4823 & 6.5236 & 11.8169 \\
Ccca (SG:68) & & 6.4828 & 6.5251 & 11.8112 \\
\hline
\multicolumn{5}{c}{optPBE-vdW} \\
\hline
Aea2 (SG:41) & & 6.6514 & 6.6854 & 12.0248 \\
Ccca (SG:68) & & 6.6515 & 6.6854 & 12.0257 \\
\hline
\multicolumn{5}{c}{optB88-vdW} \\
\hline
Aea2 (SG:41) & & 6.6130 & 6.6428 & 11.9811 \\
Ccca (SG:68) & & 6.6130 & 6.6427 & 11.9814 \\
\end{tabular}
\end{ruledtabular}
\end{table}

\section{$\mathrm{Aea2}$ and $\mathrm{Ccca}$ symmetries}
\label{sec.init_sym}

\subsection{Crystal structure}
\label{sec.crys}

As mentioned before, the AuSn$_{4}$ compound is predicted to have either Aea2~\cite{kubiak.81,kubiak.wolcyrz.85,okamoto.93,shen.kuo.20} or Ccca~\cite{zavalij.zribi.02} symmetry according to XRD measurements. 
The experimental and theoretical lattice parameters for both symmetries are provided in Tab.~\ref{tab.latt_param}.
We see that the experimental values are quite similar to each other (for reported Aea2 and Ccca symmetries), while the theoretical values are in good agreement with them (independent of the implemented vdW corrections).

\begin{figure}[!t]
\centering
\includegraphics[width=\linewidth]{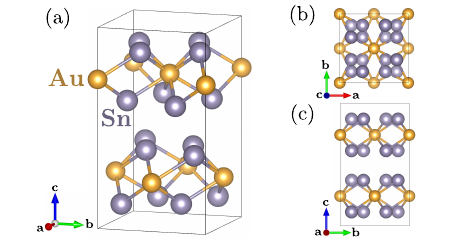}
\caption{
The general (a), top (b), and side view (c) on conventional cell of AuSn$_{4}$ crystal with Aea2 symmetry.
\label{fig.crys}
}
\end{figure}

The AuSn$_{4}$ crystal is a 2D layered compound, which consists of two AuSn$_{4}$ sheets oriented in the direction of the longest vector (Fig.~\ref{fig.crys}). While the Sn atoms form a $\sqrt{2}\times\sqrt{2}$ square-like lattice structure, the Au atoms are bonded to eight Sn atoms in an 8-coordinate geometry. The spread of the Au-Sn bond distances ranges from $2.85$~\AA\ to $2.98$~\AA. Here, the bonding between Sn to two equivalent Au atoms is reminiscent of the distorted water-like geometry. 

To study the symmetries, we optimize the structure of AuSn$_{4}$.
For example, in the absence of vdW corrections, we find the positions of atoms as follows:
\begin{itemize}
\item Aea2 symmetry (space group No.~$41$, earlier named Aba2) -- Au atom in Wyckoff position $4a$ ($0$, $0$, $0.7420$), while Sn atom in two nonequivalent Wyckoff positions $8b$ ($0.1669$, $0.1633$, $0.3715$) and ($0.6669$, $0.6633$, $0.1124$).
\item Ccca symmetry (space group No.~$68$) -- Au atom in Wyckoff position $4b$ ($0$, $\sfrac{1}{4}$, $\sfrac{3}{4}$), and Sn atom in Wyckoff position $16i$ ($0.8367$, $0.3795$, $0.4169$).
\end{itemize}
In practice, there are few free parameters describing the atoms positions.
However, these parameters are weakly influenced by the vdW corrections.

\begin{figure}[!t]
\centering
\includegraphics[width=\linewidth]{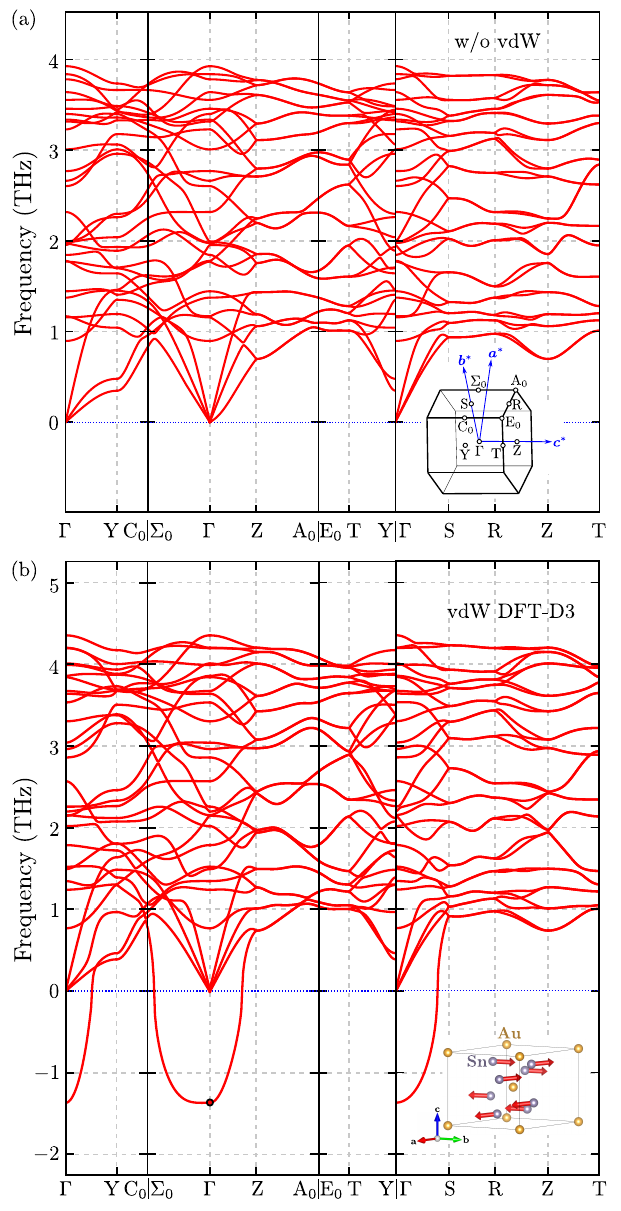}
\caption{
Phonon dispersion curves for AuSn$_{4}$ with Aea2 symmetry along the high symmetry directions of the Brillouin zone (presented on top inset).
Results obtained in (a) absence and (b) presence of van der Waals correction.
Atoms displacements in primitive cell induced by the imaginary soft mode at $\Gamma$ point (marked by black dot) are presented in bottom inset.
\label{fig.ph41}
}
\end{figure}


\begin{figure}[!t]
\centering
\includegraphics[width=0.75\linewidth]{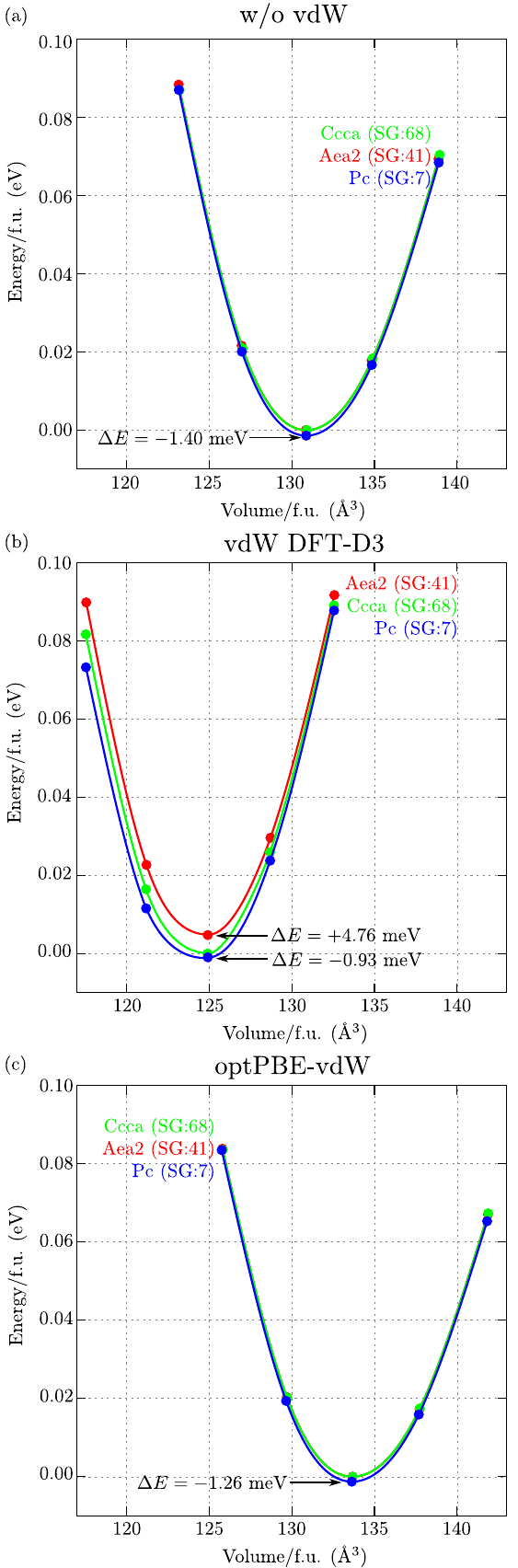}
\caption{
Dependence of the system energy on volume, for AuSn$_{4}$ with different symmetries and vdW corrections: (a) in absence of vdW correction, (b) with vdW correction within DFT-D3 method of Grimme with zero-damping function, and (c) within non-local optPBE-vdW formulation.
Color of lines correspond to the different symmetries (as labeled).
\label{fig.e_v}
}
\end{figure}

\begin{figure}[!t]
\centering
\includegraphics[width=\linewidth]{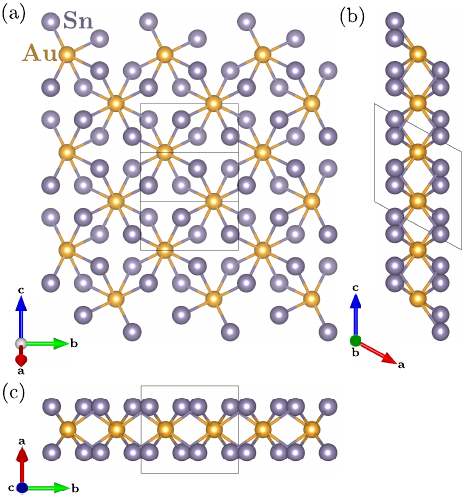}
\caption{
AuSn$_{4}$ crystal structure with Pc symmetry. 
Primitive cell is marked by the solid line.
\label{fig.crys7}
}
\end{figure}

\subsection{Dynamical study}
\label{sec.dynam}

We present the phonon dispersion curves of AuSn$_{4}$ with Aea2 symmetry on Fig.~\ref{fig.ph41}.
At first, we notice that the dispersion curves for Aea2 and Ccca symmetry (not shown) are indistinguishable along the high symmetry directions.
As we mentioned earlier, the vdW dispersion correction is added to the interatomic forces which indirectly modifies the obtained phonon dispersion curves.
In the absence of the vdW correction the phonon spectra does not exhibit imaginary soft modes and system with Aea2 (or Ccca) seems stable in regards to the positive phonon spectra [Fig.~\ref{fig.ph41}(a)].
Similar results can be obtained within the non-local vdW-DF formulation (not presented).
Contrary to this, the imaginary phonon soft modes are present when the vdW corrections are introduced within DFT-D3--like methods [Fig.~\ref{fig.ph41}(b)].
In this case, one of the optical branches become soft mode with maximal magnitude at $\Gamma$ point [marked by black dot on Fig.~\ref{fig.ph41}(b)].

Note that, the atom displacements induced by the imaginary soft mode, can be used to generate systems with new symmetry. 
In fact, this technique was successfully used recently for the stability estimation of several compounds, such as RhPb~\cite{ptok.kobialka.21}, NbReSi~\cite{basak.ptok.23} or TaReSi~\cite{ptok.24}, $A$V$_{3}$Sb$_{5}$ ($A$=K, Rb, Cs)~\cite{ptok.kobialka.22}, and Pt$_{3}$Pb$_{2}$S$_{2}$~\cite{basak.kobialka.23}. 
In our case, we find that the structure with displaced atoms after optimization possesses Pc symmetry, i.e. structure with lowered symmetry relative to the initially proposed symmetry.

In our case, the imaginary soft mode is realized at $\Gamma$ point.
This suggest the atom displacement is realized within the unit cell.
Indeed, investigation of the atom displacement induced by this soft mode [presented on the bottom inset on Fig.~\ref{fig.ph41}(b)] show that all displacements are only due to the Sn atoms.
Atoms in the neighboring Sn-layer are shifted in the same direction, alternately.
Such displacement leads to additional unit cell distortion during further relaxation which leads to the structure with lower symmetry.


\begin{figure}[!t]
\centering
\includegraphics[width=\linewidth]{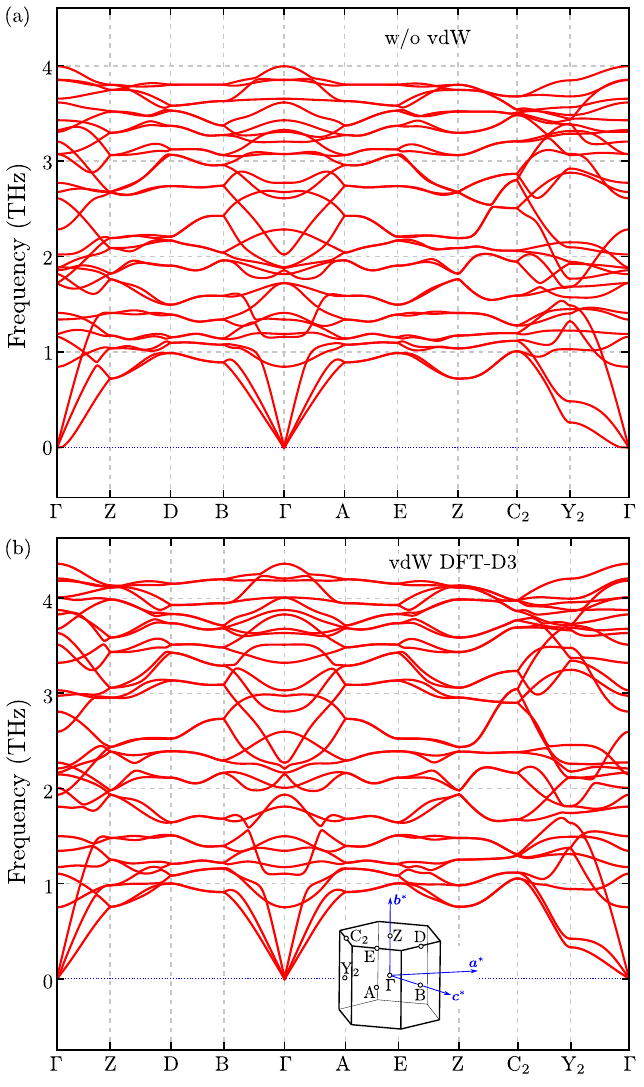}
\caption{
Phonon dispersion curves for AuSn$_{4}$ with Pc symmetry along high symmetry directions of the Brillouin zone (presented on inset).
Results obtained in (a) absence and (b) presence of van der Waals correction.
\label{fig.ph7}
}
\end{figure}

\begin{figure}[!t]
\centering
\includegraphics[width=\linewidth]{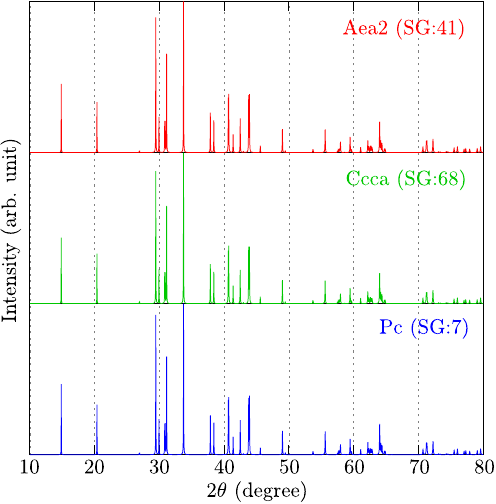}
\caption{
Theoretically obtained XRD pattern for AuSn$_{4}$ with investigated structures (as labeled).
\label{fig.xrd}
}
\end{figure}

\begin{figure*}
\centering
\includegraphics[width=0.75\linewidth]{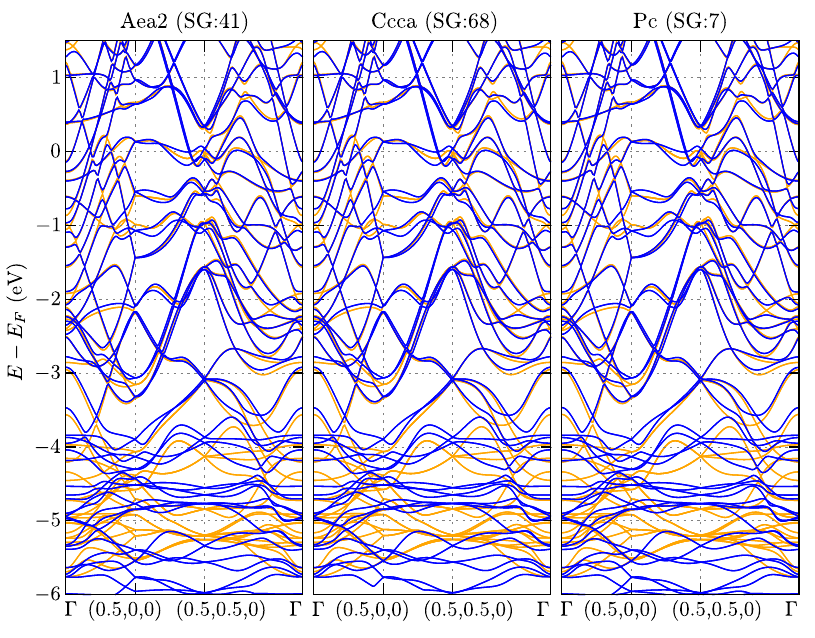}
\caption{
Comparison of the (folded) electronic band structure of AuSn$_{4}$ with different symmetries (as labeled).
The results obtained for conventional unit cell like structure in convention $a < b < c$).
The orange and blue lines correspond to the absence and presence of the spin--orbit coupling.
\label{fig.el_band}
}
\end{figure*}

\begin{figure}[!b]
\centering
\includegraphics[width=\linewidth]{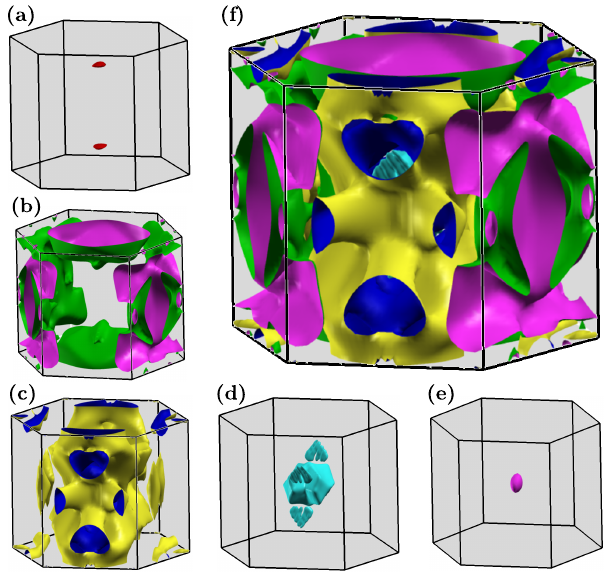}
\caption{
The separate Fermi surface pockets of AuSn$_{4}$ with Pc symmetry.
Panels from (a) to (e) present separate Fermi pockets, while (f) show full Fermi surface.
\label{fig.fermi}
}
\end{figure}

\begin{figure}[!b]
\centering
\includegraphics[width=\linewidth]{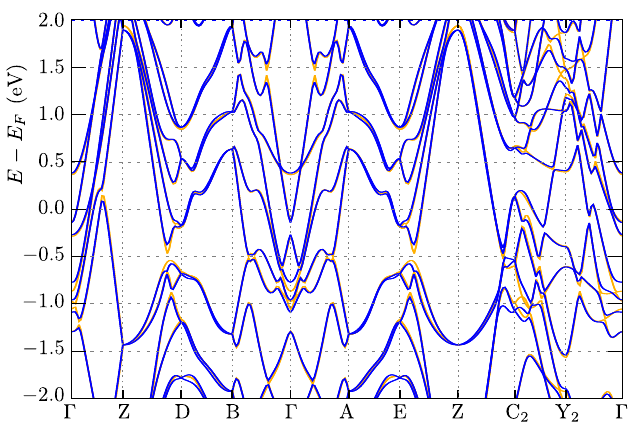}
\caption{
The electronic band structure of AuSn$_{4}$ with Pc symmetry.
Orange and blue color lines correspond to the band structure without and with spin--orbit coupling, respectively.
\label{fig.el_band_small}
}
\end{figure}

\begin{figure*}
\centering
\includegraphics[width=\linewidth]{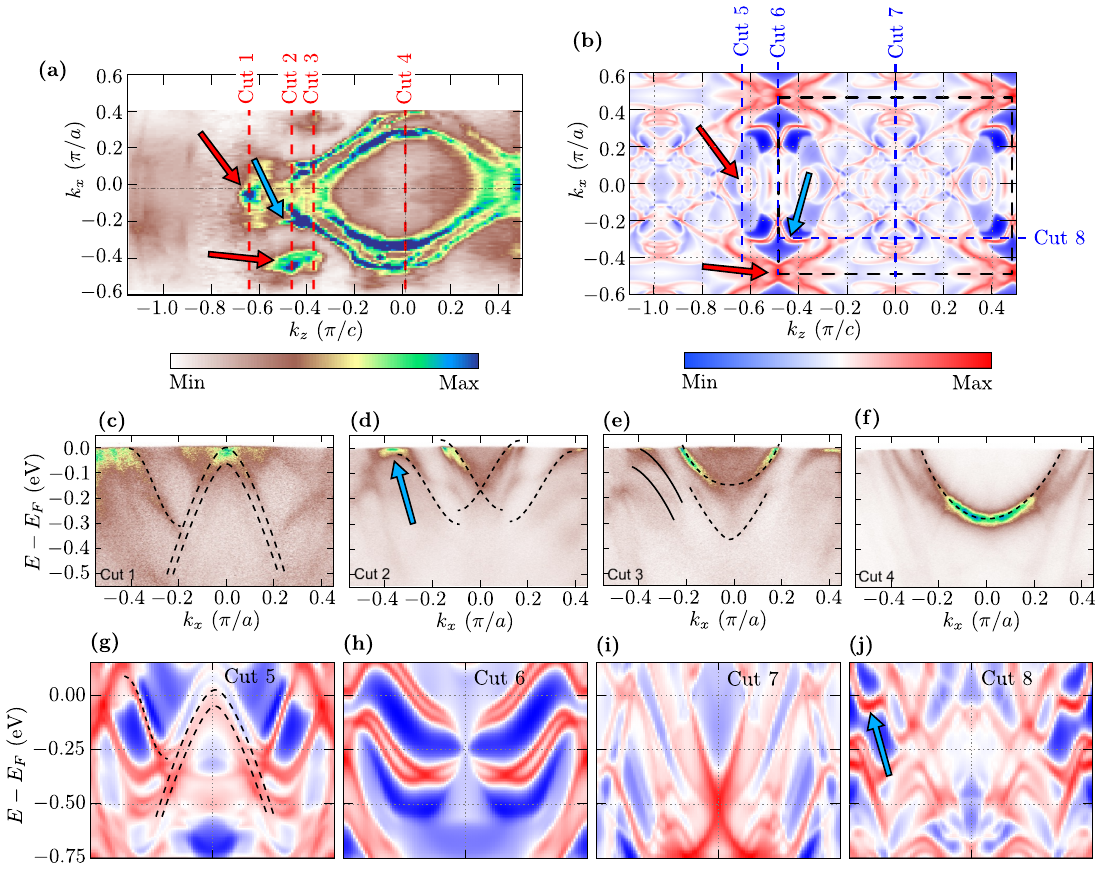}
\caption{
Comparison of the experimental ARPES spectra (a) and theoretical surface Green function (b).
Top panels present constant energy contour (for the Fermi level).
The surface Brillouin zone is marked by dashed black line on (b).
Panels (c)--(j) present spectral function for chosen cuts along the Brillouin zone, as shown in panels (a) and (b).
Experimental results presented on (a) and (c)--(f) are adopted from Ref.~\cite{herrera.wu.23}.
Here, $xz$ plane denotes the plane parallel to the layer within the AuSn$_{4}$ crystal.
\label{fig.arpes}
}
\end{figure*}

\section{$\mathrm{Pc}$ structure}
\label{sec.pc_sym}

\subsection{System stability}
\label{sec.ph7}

Comparison of the system energies with respect to the volume for Aea2, Ccca, and Pc structures are presented on Fig.~\ref{fig.e_v}.
The ``zero'' energy is set as energy for Ccca structure (green line), while points from left to right corresponds to the compressed system (with $98$\% of equilibrium volume) and the stretched system (with $102$\% of equilibrium volume).
First, in the absence of the vdW correction [Fig.~\ref{fig.e_v}(a)] or in the case of non-local vdW-DF formulation [Fig.~\ref{fig.e_v}(c)], the energies of Aea2 and Ccca structures are the same, regardless of the change in volume.
In the exclusive case with DFT-D3--like vdW correction, the Aea2 structure posses bigger energy than Ccca structure [cf.~red and gree lines on Fig.~\ref{fig.e_v}(b)].
However, independent of the presence or absence of vdW corrections, the Pc symmetry (blue line) always has smaller energy than Aea2 or Ccca structures.

Here, we should also notice that the different formulation of the vdW corrections lead to the modification of the equilibrium parameters, like e.g. lattice constant, which is apparent in Fig.~\ref{fig.e_v} as a system volume changes.
In the case of DFT-D3--like (non-local vdW-DF formulation) the volume decrease (increase) with respect to this obtained from calculations w/o vdW correction.
Finally, the system volume is close to this as reported experimentally (i.e. around $124$~\AA$^{3}$) with the DFT-D3--like vdW corrections [Fig.~\ref{fig.e_v}(b)].
This suggests that the formulation better captures the experimentally observed features.

The layered behavior of crystal structure is still observed in the obtained Pc structure [Fig.~\ref{fig.crys7}].
The lattice constants within layer are $6.6126$~\AA, and $6.6353$~\AA, which are close to experimentally reported values (see Table~\ref{tab.latt_param}).
All atoms are at Wyckoff positions $2a$, i.e., contrary to the previous case Sn atom posses four non-equivalent positions.
The main feature of the lattice shares similarity with the Aea2 and Ccca symmetries, and $\sqrt{2} \times \sqrt{2}$ square lattice of Au atoms can be recognized [see Fig.~\ref{fig.crys7}(a)].
New crystal structure can be transform to the initial one by doubling the unit cell along direction perpendicular to the layers.
Thus, the ``old'' structure can be found as (${\bm c}$, ${\bm b}$, $2{\bm a}+{\bm c}$) unit, while $\beta \approx 89^{\circ}$.
As we can see, the vdW correction lead to the additional stress within the layers of planes, and in consequence relative shift of the layers.

The phonon dispersion curves for AuSn$_{4}$ with Pc symmetry are presented on Fig.~\ref{fig.ph7}.
Contrary to the previous cases, the vdW corrections affects the phonon spectra minimally and it remains positive.
Due to the absence of the imaginary phonon soft modes the system is stable with lowered Pc symmetry.
As we mentioned in the Introduction, the Aea2 and Ccca symmetries are indistinguishable within XRD measurements~\cite{sharma.gurjar.23,karn.sharma.22}.
Unfortunately, the same problem can be encountered for the Pc structure.
Theoretically obtained XRD patterns for AuSn$_{4}$ with discussed symmetries are presented on Fig.~\ref{fig.xrd}.
In practice, all peaks in theoretical XRD are the same, and contain similar intensities, which can make it difficult to confirm the experimentally realized symmetry of AuSn$_{4}$.

\subsection{Electronic properties}
\label{sec.ele}

The electronic band structure of AuSn$_{4}$ with discussed symmetries are presented on Fig.~\ref{fig.el_band}. 
Due to the different shape of the primitive unit cells for the symmetries, we compare the folded band structure obtained for the conventional cell like structure (in convention $a < b < c$).
For simplicity we take path $\Gamma$--(0.5,0,0)--(0.5,0.5,0)--$\Gamma$.
Regardless of the realized symmetry, the obtained electronic band structure exhibits the same features.
Moreover, the properties remain similar around the Fermi level.
Similar results were reported earlier for Aea2 and Ccca symmetries in Ref.~\cite{karn.sharma.22}.

We can find several band crossings around the Fermi level, which makes the Fermi surface complex (see Fig.~\ref{fig.fermi}).
Similar to the earlier studies involving Aea2 and Ccca symmetries, the Fermi surface is formed by five pockets~\cite{herrera.wu.23, karn.sharma.22}. 
In practice, all the Fermi pockets exhibit three dimensional features.

We also studied the effect of spin--orbit coupling (SOC). 
The introduction of SOC leads to the decoupling of the bands (cf. orange and blue line in Fig.~\ref{fig.el_band_small}).
Small band decoupling is realized e.g. along A--E direction.
Nevertheless, the SOC strongly affect on the electronic band structure far below the Fermi level (below $-4$~eV, see Fig.~\ref{fig.el_band}).
Moreover, there exists multiple single-point degeneracies, which do not get lifted under SOC, and thus yield Dirac nodes in the Brillouin zone. 
This can contribute to the realization of the Dirac states in the vicinity of the Fermi level, similar to PtSn$_{4}$~\cite{wu.wang.16}.

Finally, we calculate the surface spectral function using the Green's function technique for a semi-infinite system. 
This provides us with a straightforward way to compare the numerical results with the experimental angle-resolved photoemission spectroscopy (ARPES) data (Fig.~\ref{fig.arpes}). 
For this purpose, we adopt the ARPES data from Ref.~\cite{herrera.wu.23}.
The theoretically obtained Fermi surface (FS) contour [Fig.~\ref{fig.arpes}(b)] has a shares similarity to Fermi surface observed in other related compounds, such as PdSn4~\cite{jo.wu.17} and PtSn$_{4}$~\cite{wu.wang.16,li.fu.19}.

Unfortunately, without the complete scan of the two-dimensional surface Brillouin zone, it is hard to uncover the characteristics of spectra in space [cf. Fig.~\ref{fig.arpes}(a) and Fig.~\ref{fig.arpes}(b)]. 
Nevertheless, some features of the FS contour can still be recognized. 
For example, the projection of the bulk states on the surface Brillouin zone leads to the formation of the typical peaks on the FS contours (marked by red arrows) -- one of them at the corner of the Brillouin zone.
However, we choose to analyze the FS contour cuts along directions which correspond to the experimental cuts, as indicated by the blue dashed lines in Fig.~\ref{fig.arpes}.

Turning to the experimental results, we take four line cuts along the $k_z=0$ plane -- this way we get progressively closer to the $\Gamma$ point, as we go from cut 1 to cut 4 in Fig.~\ref{fig.arpes}(a). 
These line cuts are also displayed in Fig.~\ref{fig.arpes}(c)-(f) for a better comparison. 
In cut 1, we recognize a hole-like pocket around $k_x=0$, which remains visible (around $-0.1$~eV) in cut 2 and 3. 
However, in cut 4, the hole-like band completely vanishes at the $\Gamma$ point, while an electron-like band emerges [see Fig.~\ref{fig.arpes}(f)]. 
Additionally, the Dirac-like crossing is well visible on cut 2 [Fig.~\ref{fig.arpes}(d)].

We display the theoretical line cuts, ranging from cut 5 to cut 8 in Fig.~\ref{fig.arpes}(g)--(j). 
Evidently, the theoretical line cuts reproduce the ARPES features relatively well. 
For example, the hole-like band in the experimental cuts can be recognized in cut 5 [Fig.~\ref{fig.arpes}(g)]. 
The electron-like band is well visible on cut 7, i.e. for $k_{z} = 0$ [Fig.~\ref{fig.arpes}(i)].
At the edge of the Brillouin zone [cut 6 on Fig.~\ref{fig.arpes}(h)], the Dirac-like crossing of the bands can be recognized.

Additionally, close to the edge of Brillouin zone we can recognize the surface state [marked by blue arrow on Fig.~\ref{fig.arpes}(b)].
This state is visible also along $k_{x} = const$ cut 8 [Fig.~\ref{fig.arpes}(j)].
Moreover, in the experimental spectra similar state is observed around the Fermi level [cf. blue arrows on Fig.~\ref{fig.arpes}(j) and (d)].

Finally, we should appreciate that the difference between experimental and theoretical results can be a consequence of surface oxidation.
As demonstrated in PtSn$_{4}$~\cite{dolimpio.boukhvalov.20}, the layer terminated by Sn is fully tolerant to CO molecules. 
In such case, oxidation of PtSn$_{4}$ surface occurs via the formation of a tin-oxide skin. 
Nevertheless, the valence band does not show remarkable changes after eventual oxidation.
Contrary to this, in the case of AuSn$_{4}$, oxidation can modify the topological properties~\cite{boukhvalov.dolimpio.23}.

\section{Summary}
\label{sec.sum}

In this paper, we study a AuSn$_{4}$ compound, which exhibits topological features. 
The topological features of the compound have been verified in several ways.
However, a theoretical investigation of such topological properties requires correct estimation of the crystal symmetry. 
With this in mind, we systematically analyze the AuSn$_{4}$ compound using modern {\it ab initio} techniques.
As a consequence of the layered structure of this compound, the van der Waals interaction can play an important role based on the realized symmetry.
During our investigation, we show that the lattice dynamics (and a system stability condition) strongly depends on the van der Waals correction implementation.
Thus, for the van der Waals correction within DFT-D3--like implementation, lead to the imaginary frequency soft modes, which clearly show instability of the AuSn$_{4}$ with initially suggested Aea2 or Ccca symmetries.
Thus, from the soft modes analyses we found stable Pc structure.
Unfortunately the electronic band structure or XRD pattern between described symmetry are indistinguishable, which can make confirmation of realized symmetry difficult.
Nevertheless, the structure with lowered symmetry always possess smaller energy than Aea2 or Ccca structures, independent of the van der Waals correction implementation.
According to this, the AuSn$_{4}$ can realize lower symmetry than Aea2 or Ccca in low temperatures.

\begin{acknowledgments}
Some figures in this work were rendered using {\sc Vesta}~\cite{momma.izumi.11} and {\sc XCrySDen}~\cite{kokalj.99} software. 
A.P. is grateful to Laboratoire de Physique des Solides in Orsay (CNRS, University Paris Saclay) for hospitality during a part of the work on this project. 
We kindly acknowledge the support by the National Science Centre (NCN, Poland)  under Project No.~2021/43/B/ST3/02166.
CIF files of discuss symmetries can be found in the Supplemental Material (SM)~\footnote{
See Supplemental Material at [URL will be inserted by publisher] contain CIF files with optimized structures, and additional theoretical results.}.
\end{acknowledgments}

\bibliography{biblio}

\end{document}